\documentclass[prb,preprint,amsmath,amssymb,superscriptaddress,notitlepage]{revtex4-1}
\usepackage{epsfig}
\usepackage{graphicx}
\usepackage{color}
\usepackage{bm}
\usepackage{fixme}

\begin{document}

\title{Electrically induced and detected N\'eel vector reversal in a collinear antiferromagnet}
\author{J.~Godinho}
\affiliation{Institute of Physics ASCR, v.v.i., Cukrovarnick\'a 10, 162 53, Praha 6, Czech Republic}
\affiliation{Faculty of Mathematics and Physics, Charles University in Prague, Ke Karlovu 3, 121 16 Prague 2, Czech Republic}
\author{H.~Reichlov\'a} 
\affiliation{Institute of Physics ASCR, v.v.i., Cukrovarnick\'a 10, 162 53, Praha 6, Czech Republic}
\affiliation{Institut f\"ur Festk\"orper- und Materialphysik, Technische Universit\"at Dresden, 01062 Dresden, Germany}
\author{D.~Kriegner}
\affiliation{Institute of Physics ASCR, v.v.i., Cukrovarnick\'a 10, 162 53, Praha 6, Czech Republic}
\affiliation{Max Planck Institute for Chemical Physics of Solids, 01187 Dresden, Germany}
\author{V.~Nov\'ak} 
\affiliation{Institute of Physics ASCR, v.v.i., Cukrovarnick\'a 10, 162 53, Praha 6, Czech Republic}
\author{K.~Olejn\'ik} 
\affiliation{Institute of Physics ASCR, v.v.i., Cukrovarnick\'a 10, 162 53, Praha 6, Czech Republic}
\author{Z.~Ka\v{s}par} 
\affiliation{Institute of Physics ASCR, v.v.i., Cukrovarnick\'a 10, 162 53, Praha 6, Czech Republic}
\author{Z.~\v{S}ob\'a\v{n}} 
\affiliation{Institute of Physics ASCR, v.v.i., Cukrovarnick\'a 10, 162 53, Praha 6, Czech Republic}
\author{P~Wadley}
\affiliation{School of Physics and Astronomy, University of Nottingham, Nottingham NG7 2RD, United Kingdom}
\author{R.~P.~Campion}
\affiliation{School of Physics and Astronomy, University of Nottingham, Nottingham NG7 2RD, United Kingdom}
\author{R.~M.~Otxoa}
\affiliation{Hitachi Cambridge Laboratory, Cambridge CB3 0HE, United Kingdom}
\affiliation{Donostia International Physics Center, Paseo Manuel de Lardizabal 4, Donostia-San Sebastian 20018, Spain}
\author{P.~E.~Roy}
\affiliation{Hitachi Cambridge Laboratory, Cambridge CB3 0HE, United Kingdom}
\author{J.~\v{Z}elezn\'y}
\affiliation{Institute of Physics ASCR, v.v.i., Cukrovarnick\'a 10, 162 53, Praha 6, Czech Republic} 
\author{T.~Jungwirth}
\affiliation{Institute of Physics ASCR, v.v.i., Cukrovarnick\'a 10, 162 53, Praha 6, Czech Republic} 
\affiliation{School of Physics and Astronomy, University of Nottingham, Nottingham NG7 2RD, United Kingdom}
\author{J.~Wunderlich}
\affiliation{Institute of Physics ASCR, v.v.i., Cukrovarnick\'a 10, 162 53, Praha 6, Czech Republic}
\affiliation{Hitachi Cambridge Laboratory, Cambridge CB3 0HE, United Kingdom}

\maketitle

{\bf Electrical detection of the 180$^\circ$ spin reversal, which is the basis of the operation of ferromagnetic memories\cite{Chappert2007}, is among the outstanding challenges in the research of antiferromagnetic spintronics\cite{MacDonald2011,Jungwirth2016,Baltz2018,Jungwirth2018}.  Analogous effects to the ferromagnetic giant or tunneling magnetoresistance have not yet been realized in antiferromagnetic multilayers\cite{Zelezny2018}. Anomalous Hall effect (AHE), which has been recently employed for spin reversal detection in non-collinear antiferromagnets, is limited to materials that crystalize in ferromagnetic symmetry groups\cite{Grimmer1993,Chen2014,Nakatsuji2015,Nayak2016,Zelezny2018}. Here we demonstrate electrical detection of the 180$^\circ$ N\'eel vector reversal in CuMnAs which comprises two collinear spin sublattices and belongs to an antiferromagnetic symmetry group with no net magnetic moment. We detect the spin reversal by measuring a second-order magnetotransport coefficient whose presence is allowed in systems with broken space inversion  symmetry. 
The phenomenology of the non-linear transport effect we observe in CuMnAs is consistent with a microscopic scenario combining anisotropic magneto-resistance (AMR) with a transient tilt of the  N\'eel vector due to a current-induced, staggered spin-orbit field \cite{Zelezny2014,Wadley2016,Zelezny2018}.  We use the same staggered spin-orbit field, but of a higher amplitude, for the electrical switching between reversed antiferromagnetic states which are stable and show no sign of decay over 25~hour probing times.
}

Before presenting the experimental data, we first elaborate in more detail on a microscopic mechanism that gives the seemingly counter-intuitive possibility for detecting 180$^\circ$ spin reversal in a collinear antiferromagnet comprising two chemically identical spin-sublattices. The mechanism is illustrated in Figs.~1a-c. It is based on the observation that the sites occupied by nearest-neighbor Mn atoms in CuMnAs are locally non-centrosymmetric inversion partners. This implies that electrical current induces a non-equilibrium spin-polarization with opposite sign on the two sites\cite{Zelezny2014,Wadley2016}. Simultaneously, the inversion-partner Mn sites belong to opposite spin-sublattices of the bipartite N\'eel order ground state \cite{Zelezny2014,Wadley2016}. Since the staggered current-induced polarization, and corresponding staggered effective field, are commensurate with the N\'eel order, the antiferromagnetic moments can be deflected by relatively weak currents. The electrically induced N\'eel vector deflection combined with AMR can then yield a second-order magneto-transport effect  applicable for detecting the 180$^\circ$ N\'eel vector reversal. Later in the discussion part we show that this microscopic mechanism is consistent with a general symmetry-based picture in which the spin-reversal detection by a second-order magneto-resistance is allowed in antiferromagnets ordering in magnetic point groups with broken time and space-inversion symmetry. In the next paragraph we continue by illustrating the experimental implementation of this detection technique. 

We recall that in the tetragonal lattice of CuMnAs,  the staggered field generated by a current applied in the $a-b$ plane is along the in-plane axis oriented perpendicular to the current, as highlighted in Fig.~1a\cite{Wadley2016}. Considering this geometry, we sketch in Fig.~1b a set-up for detecting the 180$^\circ$ reversal of the N\'eel vector pointing 45$^\circ$ rotated to the x-axis of the current. Here the reversal is measured by the 
longitudinal current-dependent resistance $\delta R_{xx}$. Another example of the measurement set-up is shown in Fig.~1c where we sketch the detection of the reversal of the N\'eel vector pointing along the $x$-axis via the current-dependent transverse resistance $\delta R_{xy}$. (For more details on the detection scheme see Supplementary information.)

To perform the experiment we need, apart from the readout method, also a tool allowing us to reverse the N\'eel vector in CuMnAs. For this we employ again the current-induced staggered spin-orbit field. Unlike the weaker currents applied to induce transient changes of the N\'eel vector angle during readout, for writing  we apply higher amplitude currents and the bistable 180$^\circ$ reversal is controlled by flipping the polarity of the writing current\cite{Jungwirth2016,Roy2016}.  We note that the analogous writing method was used in earlier studies of 90$^\circ$ N\'eel vector reorientation   in CuMnAs and Mn$_2$Au, controlled in this geometry by two orthogonal writing current lines and detected by the linear-response AMR\cite{Zelezny2014,Wadley2016,Olejnik2017,Bodnar2018,Meinert2017,Zhou2018}. 

Devices used in our experiments were fabricated from a 10~nm thick CuMnAs film grown by molecular beam epitaxy on a GaAs substrate\cite{Wadley2013} and protected by a 3~nm Pt layer. The sheet resistance of the stack is 100~$\Omega$. Note that the Pt cap provides additional Joule heating when the writing pulses are applied to the stack. The Joule heating assists but is not governing the deterministic, polarity-dependent switching. Further discussion  of the structure of our materials and measurements on a CuMnAs film capped with Al are presented in the Supplementary information. 

The wafers were patterned into Hall cross structures with added contacts to enable simultaneous detection of transverse and longitudinal signals, as shown on the scanning electron micrograph of the device in Fig.~2a. The longitudinal (linear-response) resistance of the structure is approximately 1~k$\Omega$. In our detection experiments,  the device is biased by a low frequency ($\omega/2\pi =143$~Hz) probing current $J_0\sin(\omega t)$ with an effective value of $J_{ac}=J_0/\sqrt2 $. We use lock-in amplifiers to measure simultaneously first harmonic (1$\omega$) and second harmonic (2$\omega$) components of the voltage signals. The former detects the linear-response AMR. The latter probes the second-order magneto-transport response  which we associate, following the mechanism in Fig.~1, with AMR combined with a periodic variation of the  current-induced staggered field and the corresponding periodic N\'eel-vector deflection (see Fig.~2b). Note that the second-order transport effects would also appear, in principle, in the zeroth harmonic voltage component. In our off-resonance experiments, however, this component is difficult to extract from the measurement noise. The second harmonic component, on the other hand, can be accurately measured by employing the homodyne detection method. For more details on our  experimental methods see Supplementary information.

Key results of our experiments are summarized in Figs.~2c,d where the plotted second-harmonic resistance is obtained by dividing the corresponding second-harmonic voltage by the probing current $J_{ac}$. In Fig.~2c we first sent a 20~ms long writing pulse $J_p$ of amplitude 11~mA (corresponding to a current density $j_p\sim 10^{7}$~A/cm$^{2}$ flowing through the CuMnAs film) along the $y$-direction  to set the N\'eel vector along the $x$-axis. We then measure for $40$~s the resulting second-harmonic transverse resistance $R_{xy}^{2\omega}$ (see Fig.~2b) with a probing current $J_{ac}= 2$~mA applied along the $x$-axis. Next we flip the polarity of the writing pulse in order to reverse the N\'eel vector and again measure $R_{xy}^{2\omega}$ with the same probing current. The sequence is repeated several times. As expected for the second-order magneto-resistance mechanism described in Fig.~1, we observe reproducible $R_{xy}^{2\omega}$ signals that are distinct for the two reversed states of the antiferromagnet. Fig.~2d shows the same type of experiments for one of the reversal sequences but with the probing performed for each state over 25 hours. The results highlight the stability of the detected 180$^\circ$ reversal signal which exhibits no sign of decay at these long probing times. 

The mechanism described in Fig.~1 suggests that we should not detect any reversal signal in $R_{xy}^{2\omega}$ if both the probing and setting currents are applied along the same direction ($x$-axis). This is because we set the N\'eel vector in this case collinear to the direction ($y$-axis) of the staggered effective field induced by the probing current and, therefore, no transverse deflection of the N\'eel vector is induced by the probing current. The picture is confirmed by the measured data shown in Fig.~3a where we apply a sequence of writing pulses along $\pm y$ and $\pm x$-directions which are indicated by red/orange and dark/light green arrows in the device sketches in the figure. In Fig. 3a, the probing current is along the $x$-axis and $R_{xy}^{2\omega}$ can only detect the reversal between N\'eel vectors set along the $x$-axis by writing current pulses along the $y$-axis (red/orange). On the other hand, $R_{xy}^{2\omega}$ is negligible for states with N\'eel vectors set along the $y$-axis by writing current pulses along the $x$-axis (dark/light green). To highlight that it is indeed the second-order magneto-resistance probing that is not effective in this geometry and not an inability in our material to set the N\'eel vector along the $y$-axis we rotate the detection setup in Fig.~3b by 90$^\circ$. When sending the probing current in the $y$-direction and measuring $R_{yx}^{2\omega}$ we can now detect the reversal between the N\'eel vector states set along the $y$-axis (by writing current pulses along the $x$-axis). Consistently, the reversal of the antiferromagnetic order between states set along the $x$-axis (by writing current pulses along the $y$-axis) is not detectable by $R_{yx}^{2\omega}$, as also seen in Fig.~3b.

Since we can write four distinct states in our device with N\'eel vectors set along $\pm x$ and $\pm y$-axes we can compare in Figs.~4a,b the second harmonic signal with the first harmonic AMR.  We again show several pulsing sequences but, unlike Fig.~2c, we now rotate the pulsing current successively in steps of 90$^\circ$ within each  sequence. The probing signals are averaged over 30~s detection time and error bars correspond to the standard deviation. Note that the larger error bars in the first-harmonic signal are typical for the longitudinal resistance in which the AMR generates only a small additional contribution (of less than 1\% in the present experiment) on top of a large isotropic resistance of the device and where the latter can show, e.g., a significant drift with temperature\cite{Wadley2016}. Still we observe a clear switching signal in $R_{xx}^{1\omega}$ which, as expected for the linear-response AMR, allows us to distinguish states with N\'eel vectors set along the $x$-axis from states set along the $y$-axis, and gives no sensitivity to the 180$^\circ$ reversal. This, in turn, is detected in the same reorientation sequence by the second-harmonic signal (e.g. $R_{xy}^{2\omega}$ for N\'eel vector reversal along the $y$-axis). We also point out that the signs of the second and first harmonic signals in Fig.~4a,b are consistent with the microscopic picture of the second order magneto-resistance originating from the combined effect of the current-induced deflection of the N\'eel vector due to the staggered spin-orbit field and the AMR.

In Fig.~4c we show the first and second-harmonic signals as a function of the amplitude of the writing current pulses. Both signals show a common threshold of the writing current and a subsequent increase with increasing current amplitude. This implies that a similar amplitude of the staggered effective field and/or similar  assisting Joule heating is required for setting any of the four measured N\'eel vector directions. 
In Fig.~4d we show the dependencies of the first and second-harmonic signals on the probing current. As expected for the linear-response transport coefficient, the first-harmonic resistance is independent of the probing current, apart from a small scatter generated by the noisy $R_{xx}^{1\omega}$ signal. In contrast, the second-harmonic resistance increases with the probing current, consistent with the second-order nature of this magneto-transport coefficient. 

In Fig.~5 we show that the studied CuMnAs film shows an easy-plane-like behavior allowing us in principle to set the N\'eel vector in any in-plane direction.  To illustrate this we apply the writing current pulses along directions rotated by $\pm45^\circ$ from the main cross axes (as shown in the inset of Fig.~5b) by biasing both legs simultaneously\cite{Olejnik2017}. The writing bias voltage is adjusted to generate again a current density $j_p\sim 10^{7}$~A/cm$^{2}$ in the cross center. Data in Figs.~5a,b are plotted for one sequence of cross-diagonal writing currents rotated successively in steps of 90$^\circ$. For this writing geometry the 90$^\circ$ N\'eel vector reorientation signal is detected in $R_{xy}^{1\omega}$, while the 180$^\circ$ reversal is probed by $R_{xx}^{2\omega}$. 

The $R_{xy}^{1\omega}$ signal in Fig.~5a shows a significant decay over the probing time of 2.5~min starting 5~s after the writing pulse. This together with the increasing signal with the increasing writing current amplitude (Fig.~4c) points to a multi-domain nature of the active region of the device. The observation is consistent with results of previous  90$^\circ$ reorientation experiments utilizing both electrical probing and x-ray magnetic linear dichroism microscopy\cite{Wadley2016,Grzybowski2017,Olejnik2017}.

Remarkably, the counterpart 180$^\circ$ reversal signal in Fig.~5b, as well as the second-harmonic  reversal signals in Figs.~2 and 3, show no decay from 5~s after the writing pulse when we initiate the electrical readout. We interpret this as follows: The first-harmonic signal measured 5~s after the pulse is already relatively low in the present experiment, corresponding to AMR of 0.08\%. Note that in other CuMnAs films, microstructures, or setting conditions we can observe two orders of magnitude larger AMR signals \cite{Kaspar2018}. Magneto-striction  is a mechanism  that can explain the relaxation of the  90$^\circ$ reorientation signal in our thin film. Because of the locking of the antiferromagnet's lattice to the substrate, the system may tend to minimize its energy by breaking into domains with the N\'eel vector randomized within a semicircle around the initial setting direction. This would diminish the  90$^\circ$ reorientation signal towards zero. 

On the other hand, the magneto-striction mechanism is even in the magnetic order parameter and, therefore, does not drive sign flips of the N\'eel vector. As a result, the randomization of the N\'eel vector is limited to the semicircle and, consequently, the 180$^\circ$ reversal signal would not drop below $2/3\pi$ times the signal corresponding to the single domain fully reversed state. From the comparison between Figs.~5a and 5b we surmize that a significant randomization within the semicircle is already completed before we initiate the readout measurement and that the remaining small changes are not observable within the experimental noise on top of the large second harmonic signal but are detectable in the weak first harmonic signal. As a result of the tendency of our antiferromagnetic structure to break into domains with N\'eel vector distributed within a semicircle around the initial setting direction, we observe the reproducible, stable easy-plane-like 180$^\circ$ reversals in the second-order magnetoresistance. 

Further details on the comparison between the first and second-harmonic signals are provided in Figs.~5c,d. First, we extended in Fig.~5c the probing time to 12 hours to highlight the stability of the second-harmonic signal in comparison to the first-harmonic signal which significantly decays  in the present structure. Consistently, we also see different characteristics of the first and second-harmonic signals when sending trains of pulses along one direction before changing the pulsing angle (Fig.~5d). In the first-harmonic signal we clearly resolve a memristive multi-level characteristics\cite{Wadley2016,Grzybowski2017,Olejnik2017} because the small changes of the readout signal due to successive pulses within the train can be resolved on top of the overall weak (strongly relaxed) 90$^\circ$ reorientation signal. On the other hand, the small memristive effect of the successive pulses is  not visible in the second harmonic signal. As a result, the 180$^\circ$ reversal signals measured from 5~s after the setting pulse are stable and independent of history. 

In the concluding paragraphs we discuss the detection of the 180$^\circ$ reversal by the second-order magneto-resistance in antiferromagnets from a general symmetry perspective. Before turning to the non-linear magneto-transport detection we first recall  limitations of the linear-response effects in antiferromagnets. AHE corresponds to the linear-response magneto-resistance, $E_i=\rho_{ij}^{odd}(\vec{O})\,j_j$, that is odd under time reversal $T$, i.e., $E_i=-T\rho_{ij}^{odd}(\vec{O})\,j_j=-\rho_{ij}^{odd}(-\vec{O})\,j_j$. Here $\vec{E}$ is the electric field, $T\rho_{ij}^{odd}$ labels the time-reversal operation on the resistivity tensor, $\vec{j}$ is the current density, and $\vec{O}$ is the magnetic order parameter vector that breaks $T$ symmetry of the system. 
In antiferromagnets, AHE is allowed by symmetry only in a subset  of the 122 magnetic point groups. These are the antiferromagnets that order in one of the 31 ferromagnetic symmetry point groups, i.e., can develop a net magnetic moment along some directions without changing the symmetry of the magnetic lattice \cite{Grimmer1993}. Consistent with this symmetry argument, non-collinear weak-moment antiferromagnets Mn$_3$Ir,  Mn$_3$Sb, or Mn$_3$Ge have been recently identified to host the AHE\cite{Chen2014,Nakatsuji2015,Nayak2016}. 

The antiferromagnetic lattice of CuMnAs has a broken $T$ symmetry in its magnetic point group. However, it is an example of an antiferromagnet that does not belong to one of the ferromagnetic symmetry point groups. AHE is, therefore, excluded despite the broken  $T$ symmetry. Namely, it is the combined $PT$ symmetry of CuMnAs, where $P$ is the space inversion, which makes the AHE vanish in this antiferromagnet. We can see it from the above linear response equation. Here space inversion flips sign of both the electric field and current. This implies that applying space and time inversion to the linear-response transport equation gives  $\rho_{ij}^{odd}=-PT\rho_{ij}^{odd}$. On the other hand, the $PT$ symmetry of CuMnAs and the Neumann's principle, linking the symmetries of a crystal to its physical properties,  impose that $\rho_{ij}^{odd}=PT\rho_{ij}^{odd}$. The two conditions than yield $\rho_{ij}^{odd}\equiv0$ by symmetry.

AMR is a complementary linear-response effect allowing to detect the direction of the order parameter in magnetic films. AMR is in principle present in any of the 31 ferromagnetic symmetry point groups and also in any of the  remaining 91 symmetry point groups of "true" antiferromagnets that do not allow for a net magnetic moment without changing the symmetry of the magnetic crystal. Within these 91 point groups, AMR has been detected in CuMnAs, as well as in FeRh, MnTe, or Mn$_2$Au that all host a collinear bipartite, fully compensated N\'eel order\cite{Marti2014,Kriegner2016,Wadley2016,Bodnar2018,Meinert2017,Zhou2018}.  However, AMR corresponds to the linear-response magneto-resistance coefficient that is even under time reversal, $\rho_{ij}^{even}(\vec{O})=\rho_{ij}^{even}(-\vec{O})$, i.e., gives the same electrical signal when reversing spins by 180$^\circ$. Note that the 91 groups allowing for no net moment in the point group split in 32 $T$-symmetric point groups of antiferromagnets that are invariant under anti-translations ($T$ combined with translation) and in the remaining 59 antiferromagnetic point groups with broken $T$-symmetry.  

By measuring the second-order magneto-transport coefficient we can extend the detection of the 180$^\circ$ spin reversal from antiferromagnets  within the ferromagnetic point groups  to the larger family of antiferromagnetic point groups with broken $T$-symmetry and no net moment allowed in the point group. There is an additional symmetry condition required for the presence of the second-order magneto-transport coefficient which is the broken $P$ symmetry in the antiferromagnetic lattice. This can be seen by applying the $P$ operation on the second-order transport equation (odd under $T$ reversal),  $E_i=\xi_{ijk}^{odd}\,j_jj_k$, and recalling that $P$ flips sign of both the electric field and current. This implies that, $\xi_{ijk}^{odd}=-P\xi_{ijk}^{odd}$, which allows for a non-zero  $\xi_{ijk}^{odd}$ only if $P$ is broken. 

As seen from Fig.~1a, CuMnAs is one example from the 59 antiferromagnetic point groups with broken $T$ symmetry that has also broken $P$ symmetry in the magnetic crystal. In general, 48 out of the 59 antiferromagnetic point groups and 21 out of the 32 ferromagnetic point groups have broken $P$ symmetry which makes the second-order detection method of the 180$^\circ$ spin reversal broadly applicable in antiferromagnets.

Symmetry arguments are the basis for analyzing whether a given effect can in principle exist in a certain class of materials. Its magnitude, on the other hand, is determined by the microscopic origin of the effect. Remarkably, the same combined $PT$ symmetry in CuMnAs, which excluded the AHE in this material, allows for the specific microscopic mechanism of the second-order magnetoresistance that combines current-induced deflection of the N\'eel vector with AMR. While our experiments are qualitatively compatible with this scenario,
other microscopic mechanism can contribute in CuMnAs or can govern the second-order magneto-transport detection of the 180$^\circ$ spin reversal in other antiferromagnets with broken time and space-inversion symmetries. 

\subsection*{Acknowledgment}
The authors acknowledge the EU FET Open RIA Grant No. 766566, the Grant Agency of the Czech Republic Grant No. 14-37427G, the Ministry of Education of the Czech Republic Grant No. LM2015087 and LNSM-LNSpin, and the ERC Synergy Grant No. 610115.

\section{Experimental techniques}
\subsection{Electrically induced N\'eel vector deflection combined with AMR} 
We first describe in more detail the microscopic mechanism of the second order magnetoresistance in which electrically induced N\'eel vector deflection is combined with AMR. The current-induced staggered spin-polarization (Fig.~S1a) generates spin-orbit fields $\bold{B_{SO}^{A}}$ and $\bold{B_{SO}^{B}}$, with $\bold{B_{SO}^{B}}= -\bold{B_{SO}^{A}}$,  acting on the corresponding sublattice magnetizations  $\bold{M_{A}}$ and $\bold{M_{B}}$ of the bipartite antiferromagnet CuMnAs.  $\bold{B_{SO}^{A, B}}$ are oriented perpendicular to the applied current direction and their magnitude is proportional to the applied current density $j$. In equilibrium,  $\bold{M_{A}}= -\bold{M_{B}}$, so that the corresponding spin-orbit torques $\bold{B_{SO}^{A}}\times \bold{M_{A}}=\bold{B_{SO}^{B}}\times \bold{M_{B}}$ cant the sublattice magnetisations from their antiparallel equilibrium orientation. The resulting exchange torques then rotate the sublattice magnetisations $\bold{M_{A, B}}$ within the basal plane of CuMnAs towards the direction of the spin-orbit fields $\bold{B_{SO}^{A,B}}$. Our detection method is based on the fact that the spin-orbit-torques and the resulting exchange torques flip their signs when the sublattice magnetisations reverse and therefore deflect the reversed N\'eel vector in the opposite direction (see Figs. S1b,c). This combined with AMR makes the second order magneto-resistance, in general,  unequal for the reversed states and allows for the electrical detection of the N\'eel vector reversal. 

At high-amplitude setting current pulses the antiferromagnetic moments are aligned with the direction of the current-induced spin-orbit fields \cite{Zelezny2014, Wadley2016}. 
At low probing currents (weak spin-orbit fields relative to anisotropy fields), the antiferromagnetic moments are only deflected by a small angle $\delta\varphi$ proportional to the magnitude of the current induced spin-orbit fields. This combined with the AMR results in a second-order magneto-transport effect and a corresponding resistance variation, $\delta R_{ij}$, that depends linearly on the reading current. To describe the $\varphi$-dependence of $\delta R_{ij}$ we first recall the angular dependence of the linear-response AMR. Assuming that AMR in CuMnAs is dominated by the non-crystalline component, the longitudinal AMR is given by $R_{xx}  = R_{0}+\Delta_{AMR} \cdot \cos(2\varphi)$ and the transverse AMR by  $R_{xy}= \Delta_{AMR} \cdot \sin(2\varphi)$, with $\Delta_{AMR} = \frac{1}{2}[R_{xx}(\bold{M_{A,B}} \parallel \bold{j})-R_{xx}(\bold{M_{A,B}}\perp \bold{j})]$.

Fig. S1b shows a scenario where in one panel the equilibrium N\'eel vector is set at an angle $\varphi = 45^{\circ}$  from the $x$-axis of the reading current while in the other panel the equilibrium N\'eel vector is reversed. When the current $j$ is applied, the antiferromagnetic moments  are deflected clockwise by   $-\delta\varphi$ or counter-clockwise by $+\delta\varphi$ depending on the equilibrium N\'eel vector direction. The longitudinal resistance of CuMnAs then decreases or increases by  $\delta R_{xx}$ due to the longitudinal AMR.  
In Fig. S1c, we sketch the scenario where the N\'eel vector is aligned with the $x$-axis of the reading current. In this configuration, the current induced N\'eel vector deflection results in the transversal resistance  variation $\pm\delta R_{xy}$, depending on the direction of the   N\'eel vector.  Since $\delta R_{xx}$ and $\delta R_{xy}$ are current depend, we call them \textit{nonlinear} AMR contributions in contrast to the current independent $R_{xx}$ and $R_{xy}$ which we call   \textit{linear} AMR contributions.


The easy plane magnetic anisotropy of our CuMnAs crossbar devices enabled us to set the N\'eel vector along a series of different in-plane directions (we measured 8 directions). With this we could perform extensive consistency checks between the signs of the linear and nonlinear AMR contributions measured in both longitudinal and transverse geometries. The results are in full agreement with the scenario of the second-order magnetoresistance that combines the current-induced N\'eel vector deflection with the AMR. We note that these consistency checks did not require the knowledge of the sign of the staggered current induced spin-orbit field on a given spin-subblatice for a given current direction. This is  because in our measurements of $\delta R_{xx}$ and $\delta R_{xy}$, the sign enters twice: first, when set the  N\'eel vector direction by the staggered spin-orbit field and, second, when we detect the N\'eel vector direction via the staggered spin-orbit field deflection of the N\'eel vector. 

\subsection{Detection of the nonlinear AMR} 
In order to separate the linear and nonlinear AMR contributions, we apply an alternating probing current $J_{0}\sin(\omega t)$ along the $x$-axis (corresponding to a low current density $\sim 1\times 10^{6}\text{A/cm}^{2}$) of frequency $\omega/2\pi=143$~Hz.  At such a quasi-static condition, the deflection angle and the corresponding longitudinal and transversal resistance variation follow directly the alternating current without phase-shift, so that

 $\delta R_{xx}(\varphi, t)  \sim 2 J_{0} \cdot \Delta_{AMR} \cdot \cos(\varphi) \cdot \sin(2\varphi) \cdot  \sin(\omega t)$ and 
 
 $\delta R_{xy}(\varphi, t)  \sim -2J_{0} \cdot \Delta_{AMR} \cdot \cos(\varphi) \cdot \cos(2\varphi) \cdot  \sin(\omega t)$.

\noindent Since both ac-current and device resistance oscillate at the same frequency $\omega$, Ohm's law yields, $\delta V_{xx}= \delta R_{xx}(\varphi, t)\cdot J_{ac}(t) \sim J_{0} \cdot \Delta_{AMR} \cdot \cos(\varphi) \cdot \sin(2\varphi) \cdot  (1+\sin(2\omega t- 90^{\circ}))$ and $\delta V_{xy}= \delta R_{xy}(\varphi, t)\cdot J_{ac}(t) \sim -J_{0} \cdot \Delta_{AMR} \cdot \cos(\varphi) \cdot \cos(2\varphi) \cdot  (1+\sin(2\omega t-90^{\circ}))$.
Therefore, the nonlinear AMR appears only as a time-independent constant voltage and as a second harmonic voltage signal oscillating at twice of the alternating reading current frequency. 

In our experiments we use lock-in amplifiers to measure simultaneously longitudinal and transversal voltage signals at the current frequency $\omega$ (first harmonic signals $V_{xx}^{1\omega} $ and $ V_{xy}^{1\omega}$)  and at twice of the current frequency $2\omega$ (second harmonic signals $V_{xx}^{2\omega} $ and $ V_{xy}^{2\omega}$). The  first harmonic signals contain only the linear AMR responses since the contributions from the nonlinear AMR average out to zero. From the second harmonics signal we can exclude contributions from the Joule heating since they do not depend on the N\'eel vector orientation and a possible contribution from the magneto-thermopower is an even function under N\'eel vector reversal and also small in our symmetric devices. Contributions from the anomalous Nernst effect do not appear in antiferromagnetic CuMnAs for the same symmetry reason ($PT$-symmetry) as discussed in the main text in the context of the absence of the anomalous Hall effect. We therefore  can assign linear and non-linear AMR to the measured signals as

$R_{xx}^{1\omega}(\varphi) = \text{Re}(V_{xx}^{1\omega})(\Delta \phi = 0^{\circ})/J_0 = R_{0}+\Delta_{AMR} \cdot \cos(2\varphi)$,

$R_{xy}^{1\omega}(\varphi) =\text{Re}(V_{xy}^{1\omega})(\Delta \phi = 0^{\circ})/J_0 = \Delta_{AMR} \cdot \sin(2\varphi)$,
 
 $R_{xx}^{2\omega}(\varphi)  = \text{Re}(V_{xx}^{2\omega})(\Delta \phi = -90^{\circ})/J_0 \sim \Delta_{AMR} \cdot \cos(\varphi) \cdot \sin(2\varphi)$, and
 
 $R_{xy}^{2\omega}(\varphi)  = \text{Re}(V_{xy}^{2\omega})(\Delta \phi = -90^{\circ})/J_0 \sim -\Delta_{AMR} \cdot \cos(\varphi) \cdot \cos(2\varphi)$, 

\noindent where $\text{Re}(V)$ is the part of the measured signal detected by the lock-in amplifiers which oscillates delayed by the phase-shift $\Delta \phi$ with respect to the reading current.  

\section{Device fabrication and material characterization}

Devices used for our experiments were fabricated from an epitaxial 10~nm thick tetragonal CuMnAs film grown by molecular beam epitaxy on a GaAs substrate \cite{Wadley2013} and covered in-situ by a  Pt layer of a nominal thickness of 3~nm. Additionally, a nominal 2.5~nm thick Al  layer, which almost fully oxidizes when exposed to air, was deposited on top of Pt in order to protect the film against oxidation. 

Several devices of different sizes were prepared showing qualitatively the same results. Before patterning, the CuMnAs film was measured by superconducting quantum interference device (SQUID) magnetometry to exclude any ferromagnetic impurities, uncompensated moments, or proximity polarisation in Pt. The data are shown in Fig.~S2a. 

X-Ray diffraction (XRD) was employed to confirm the quality and thickness of the layers.  Within the error-bars, the measured CuMnAs thickness corresponds to the nominal value of 10~nm and the measured 3.6~nm thickness of Pt is also close to the nominal value. The measured Al cap thickness was found to be around 4~nm, i.e., slightly thicker than the nominal value of 2.5 nm of the deposited Al layer. We explain this by the oxidation of the Al cap. The measurements are shown in Fig.~S2b.

Wafers were patterned into Hall cross devices, as shown in Fig.~2a of the main text, defined by electron beam lithography and patterned by argon plasma etch using HSQ resist mask which was removed afterwards. Electrical contacts to the sample were defined by e-beam lithography, evaporation of  Cr(5 nm)/Au(80 nm) bi-layer and followed by a lift-off  process.

\section{Effect of capping layers on switching properties of the devices}
To evaluate the effect of the 3~nm Pt layer on top of the 10~nm CuMnAs layer, a reference film was grown simultaneously by masking part of the wafer during Pt evaporation. Fig.~S3 shows the bipolar switching characteristics  of a  4-contact cross-bar device with $10~ \mu$m wide bars  patterned from the reference CuMnAs/AlOx film without the Pt-layer. Here we measured the transverse second-harmonic resistance  $R_{xy}^{2\omega}$  as a response to the probing ac-current of effective value $J_{ac} = J_0/\sqrt2  = 1$~mA ($ j_{ac}  \sim1\times 10^{6}$~A/cm$^{2}$) applied along the $x$-axis   after 20~ms long, 9~mA writing pulses ($ j_{ac} \sim9\times 10^{6}$~A/cm$^{2}$) applied along the $y$-axis. The measured  $R_{xy}^{2\omega}$   shows again the expected dependence of the second harmonics signal on the polarity of the setting current pulses corresponding to reversed N\'eel vector states. Note that in this reference sample, setting current pulses of a $\sim 30 \%$ higher current density were required. We assign the difference in required switching current densities to the difference in Joule heating  between the devices patterned from the CuMnAs/Pt/AlOx film and the devices patterned from the CuMnAs/AlOx film without Pt.

The total sheet resistance $R_{T}$ of the CuMnAs(10nm)/Pt(3nm)/AlOx stack is $\sim100~\Omega$, which is approximately $4\times$ lower than the sheet resistance of the refernce CuMnAs(10nm)/AlOx film. Therefore, in the stack containing the Pt layer, only 1/4-th of the total applied current flows through the CuMnAs layer and 3/4-th of the current flow through the highly conductive Pt layer, which increases the sample temperature during the setting current pulse and facilitates the current induced switching. Note that the Joule heating in the film containing Pt at the same current density in the CuMnAs layer is about $4\times$ larger than in  the reference CuMnAs/AlOx film, since $R_{Pt}\cdot I_{Pt}^{2}+ R_{CuMnAs}\cdot I_{CuMnAs}^{2} = 1/3 R_{CuMnAs} \cdot (3 \cdot I_{CuMnAs})^{2}+ R_{CuMnAs}\cdot I_{CuMnAs}^{2} = 4 R_{CuMnAs}\cdot  I_{CuMnAs}^{2}$.

Apart from Joule heating, an additional spin-orbit torque generated by the current flowing through the CuMnAs/Pt interface  could be considered to affect  magnetisation dynamics \cite{Zelezny2014,Jungwirth2016}. This torque can originate from the spin Hall effect in Pt or from the inverse spin galvanic (Edelstein) effect at the CuMnAs/Pt interface. Both effects would result in a  \textit{non-staggered} interfacial spin-polarisation $\bold{p}$ oriented along the $y$-axis when the current flows along the $x$-axis. In this case, the antidamping-like torque, which is driven by the sub-lattice magnetisation dependent  \textit{staggered} antidamping fields, $\sim \bold{p}\times \bold{M_{A}}$ and $\sim \bold{p}\times \bold{M_{B}}= - \bold{p}\times \bold{M_{A}}$, can, in principle, efficiently act on the antiferromagnetic state \cite{Zelezny2014,Jungwirth2016}. However, in case of CuMnAs, this interfacial spin-orbit torque remains inefficient. It cants the sub-lattice magnetisations towards the in-plane orientation perpendicular to the applied current direction and the resulting exchange torques would then trigger N\'eel vector reorientation towards the out-of-plane direction. This is inefficient, however, due to the strong, out-of-plane hard-axis anisotropy in tetragonal CuMnAs. 


\bibliographystyle{naturemag}

\newpage
\begin{figure}[h]
\hspace*{-0cm}\includegraphics[width=\textwidth]{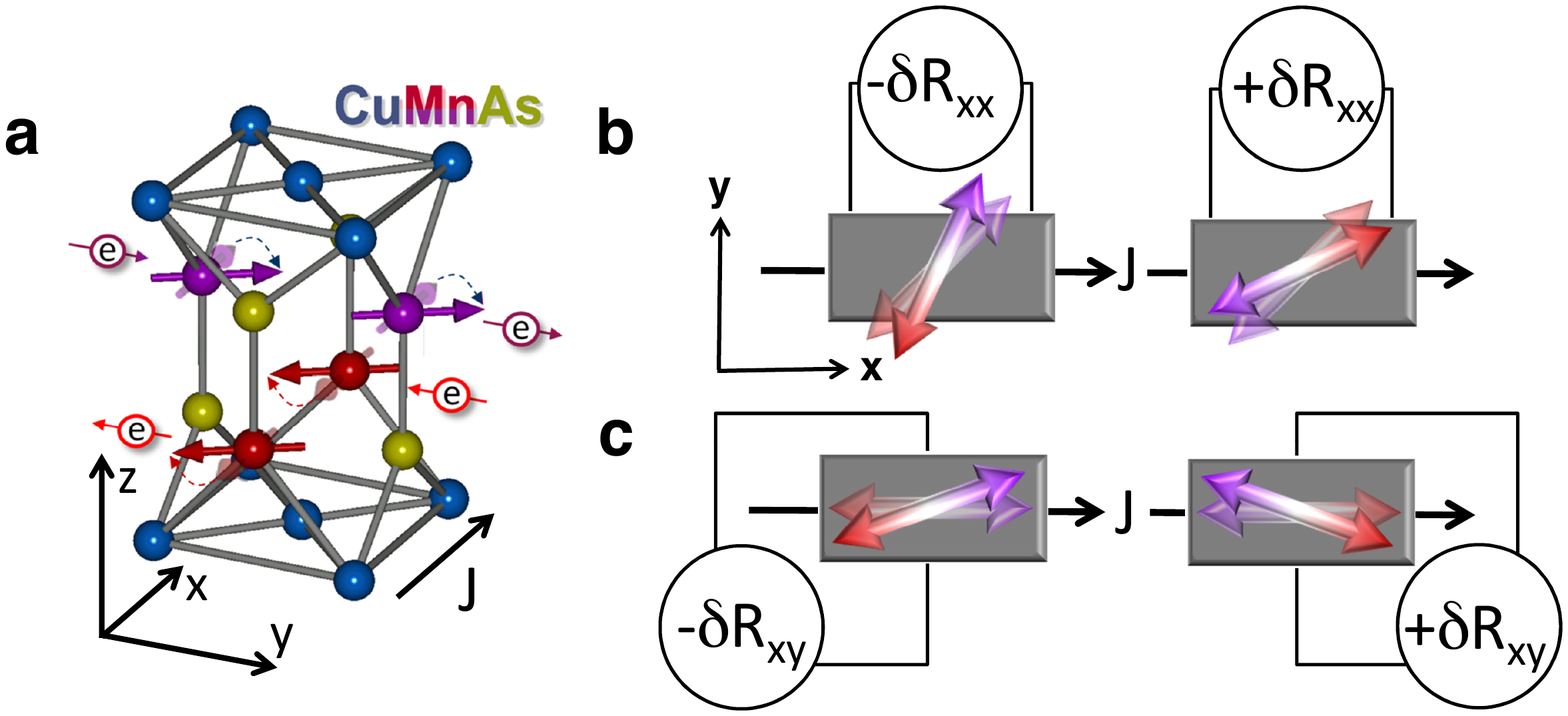}
\caption
{{\bf Schematics of the microscopic mechanism of the second-order magneto-resistance in antiferromagnetic CuMnAs.} {\bf a,} Probing current (black arrow) generates staggered non-equilibrium spin polarization (red and purple electron symbols with arrows) that causes transient deflection of the antiferromagnetic moments (thick red and purple arrows on Mn sites).  {\bf b,} $180^\circ$ reversal of the N\'eel order probed by the current-dependent resistance $\delta R_{xx}$, associated with the electrically induced deflection of antiferromagnetic moments (double-arrows) combined with AMR,  for equilibrium antiferromagnetic moments (semi-transparent double-arrows) aligned at an angle $45^\circ$ from $x$-axis of the probing current. {\bf c,} Same as {\bf b,} for $\delta R_{xy}$ and equilibrium antiferromagnetic moments aligned with $x$-axis.}
\end{figure}

\begin{figure}[h]
\hspace*{-0cm}\includegraphics[width=\textwidth]{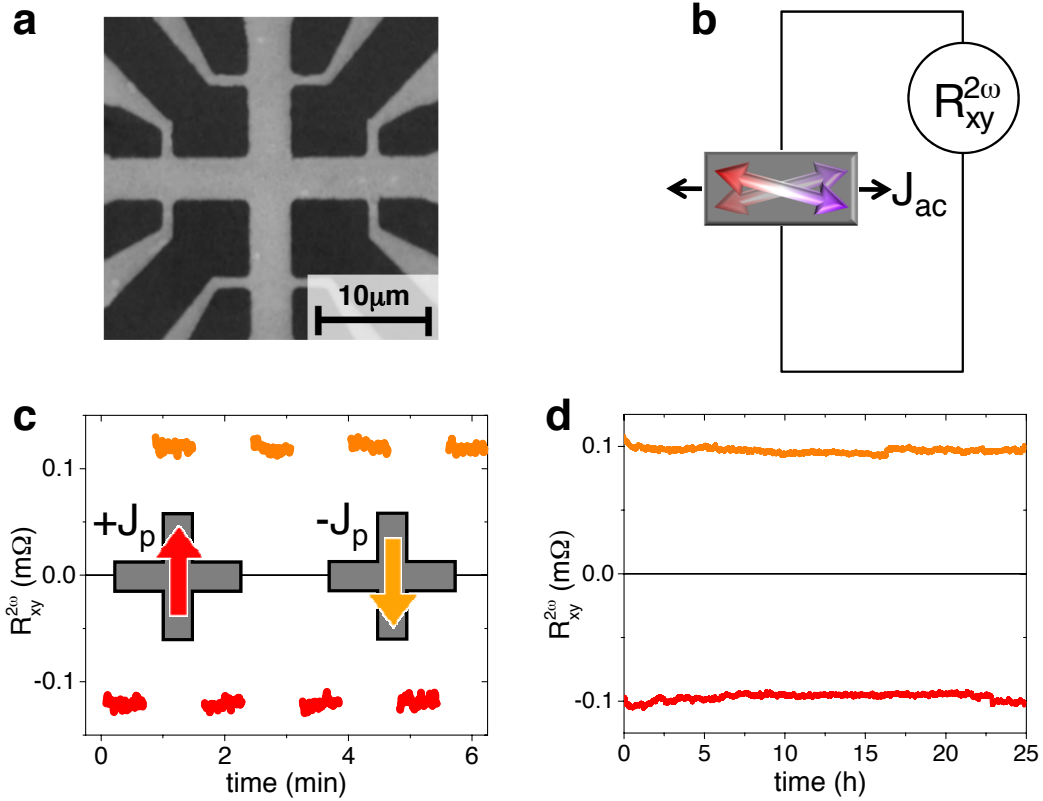}
\caption
{{\bf Electrical detection of $180^\circ$ reversal of the N\'eel order in an antiferromagnet CuMnAs.} {\bf a,} Scanning electron micrograph of the cross-bar device with contacts allowing to measure longitudinal and transverse resistances along $x$ and $y$-axes. {\bf b,} Measurement set up with a probing ac current along $x$-axis and second-harmonic voltage detected along $y$-axis, giving $R_{xy}^{2\omega}$. {\bf c,} 20~ms long pulses of the writing current  $J_p$=11~mA ($j_p\sim 10^{7}$~A/cm$^{2}$ in CuMnAs) along the $\pm y$-direction (red/yellow arrows) are applied to set the N\'eel vector along the $\pm x$-axis.  Second-harmonic transverse resistance $R_{xy}^{2\omega}$ is measured with a probing current $J_{ac} =  2$~mA  ($j_{ac}\sim 10^{6}$~A/cm$^{2}$) applied along the $x$-axis. {\bf d,} Same as {\bf c,} for one writing pulse along $+y$-axis and one subsequent pulse along $-y$-axis and 25 hour measurement of the stability of the second-harmonic probing signal.
}
\end{figure}

\begin{figure}[h]
\hspace*{-0cm}\includegraphics[width=\textwidth]{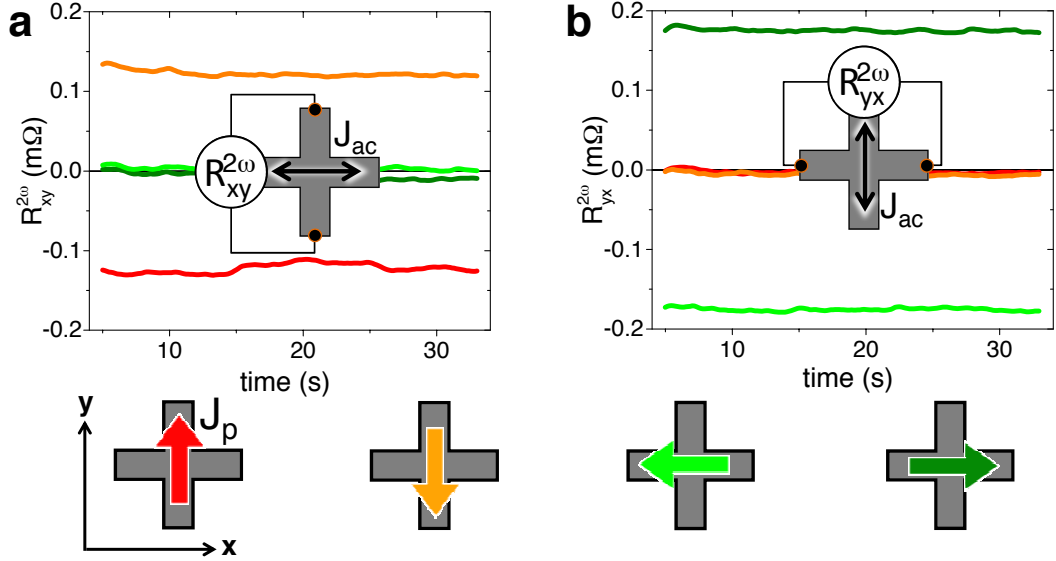}
\caption
{{\bf Symmetry of the second-harmonic signals.} {\bf a,} $R_{xy}^{2\omega}$ readout for a sequence of writing pulses along $+y$, $+x$, $-y$, $-x$ directions. {\bf b,} Same as {\bf a,} with $R_{yx}^{2\omega}$ readout (probing current along $y$-axis). Readout measurements in panels {\bf a,b}  start 5~s after the writing pulse.}
\end{figure}

\begin{figure}[h]
\hspace*{-0cm}\includegraphics[width=\textwidth]{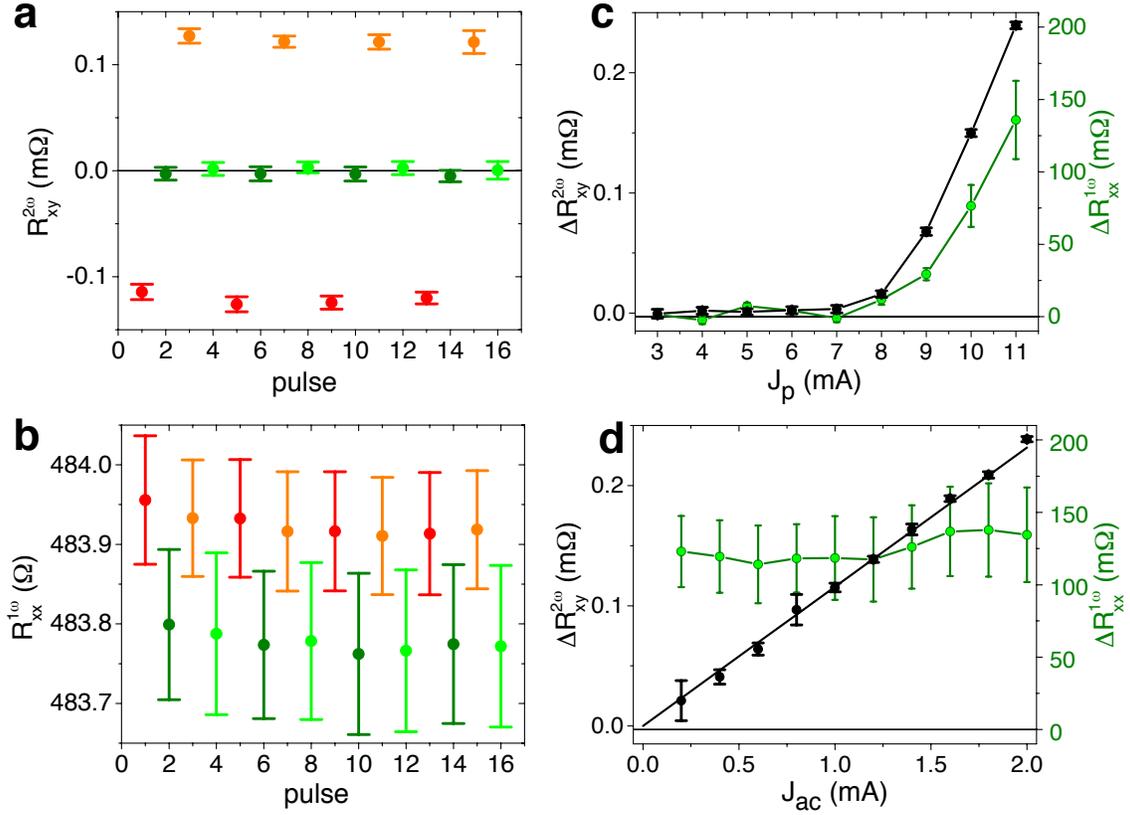}
\caption
{{\bf Comparison of second and first-harmonic signals.} {\bf a,} Second-harmonic $R_{xy}^{2\omega}$ signal measured for four sequences of writing pulses along $+y$, $+x$, $-y$, $-x$ directions. {\bf b,} Same writing sequences as in {\bf a,} probed with the first-harmonic $R_{xx}^{1\omega}$. {\bf c,} First and second-harmonic signals measured as a function of the amplitude of the writing current pulse. {\bf d,} Dependencies of the first and second harmonic signals on the probing current $J_{ac}$.  Probing signals are averaged over 30~s detection time starting 5~s after the writing pulse and error bars in {\bf a-d} correspond to the standard deviation.
}
\end{figure}

\begin{figure}[h]
\hspace*{-0cm}\includegraphics[width=\textwidth]{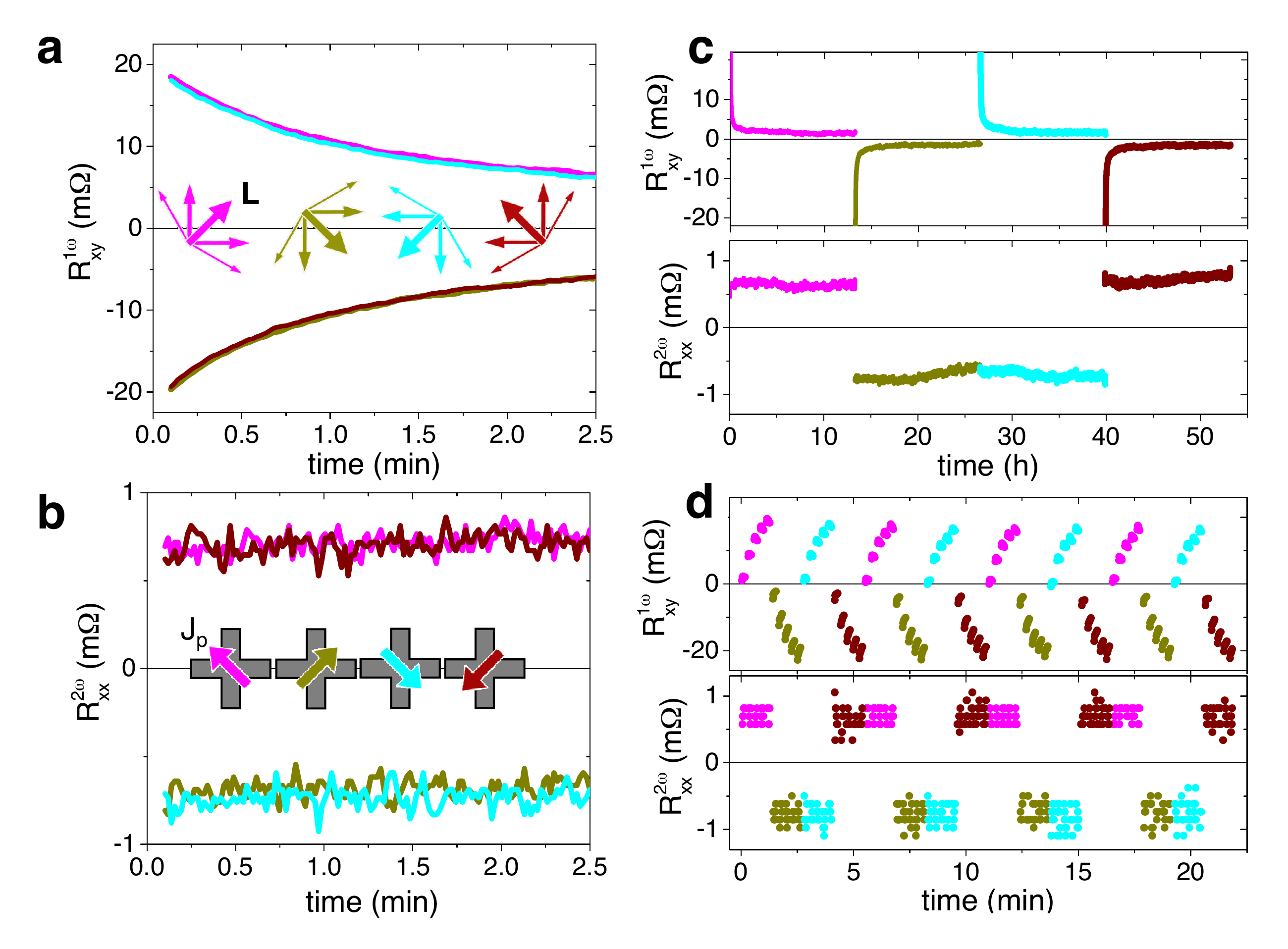}
\caption
{{\bf Time-dependence of first and second-harmonic signals.} {\bf a,b} First-harmonic $R_{xy}^{1\omega}$ and second-harmonic $R_{xx}^{2\omega}$ detection for a sequence of writing current pulses along directions rotated by $\pm45^\circ$ from the main cross axes with $j_p\sim 10^{7}$~A/cm$^{2}$ in the cross center. {\bf c} same as {\bf a,b}  for  12 hour probing measurements after each writing pulse. {\bf d,} First and second-harmonic signals measured when sending trains of five writing pulses along one direction before changing the pulsing angle. In all panels, probing starts 5~s after the writing pulse.}
\end{figure}

\begin{figure}[h!]
\hspace*{-0cm}\includegraphics[width=.8\textwidth]{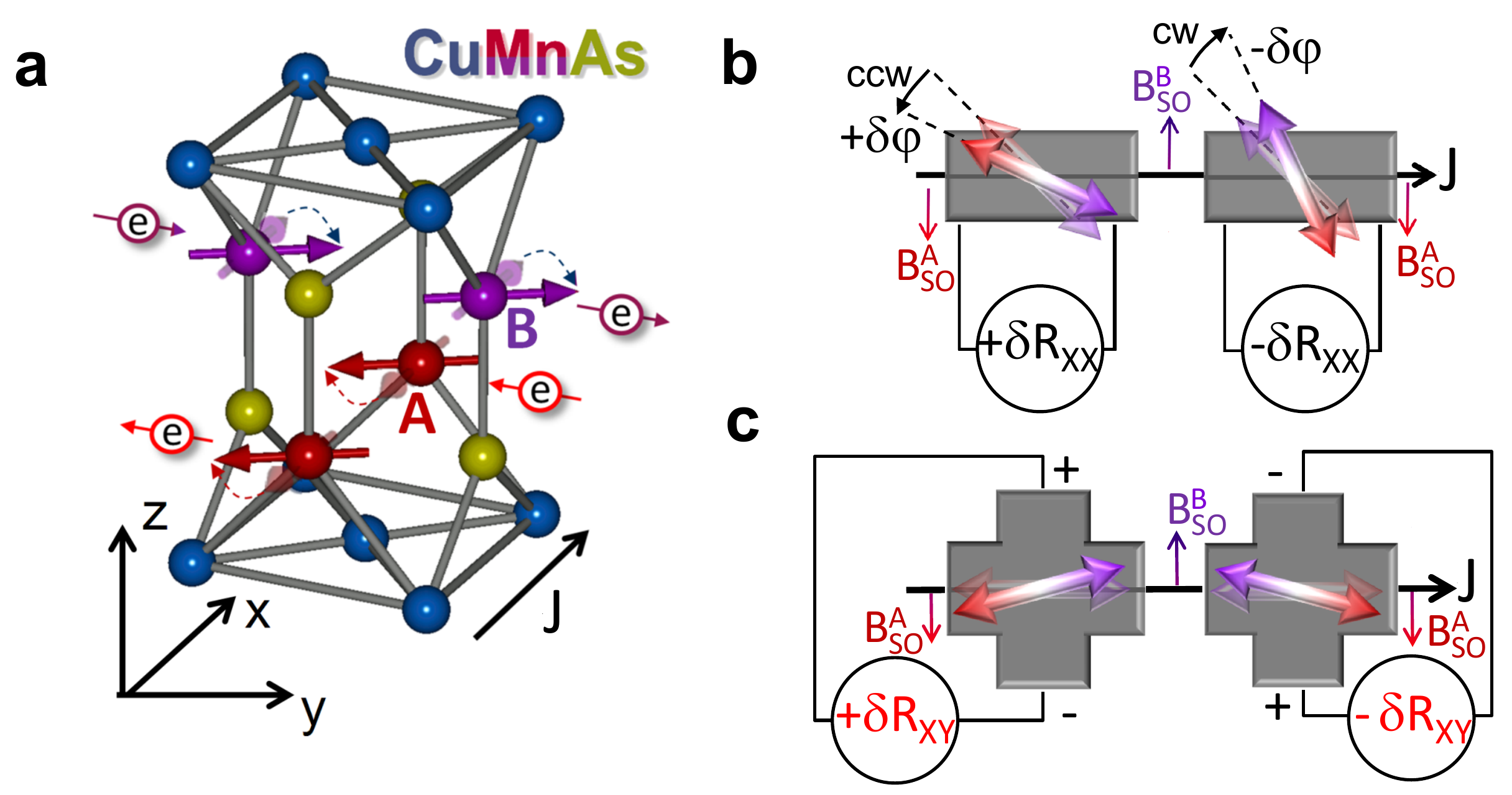}
\caption{(a) Antiferromagnetic CuMnAs with collinear spin-sublattices A and B. The two sites occupied by nearest neighbour Mn atoms are locally non-centrosymmetric inversion partners and belong to opposite spin-sublattices of the bipartite N\'eel order ground state. When biased by a charge-current $J$, a staggered spin-polarisation perpendicular to the current direction is generated with opposite sign on the two sites. The antiferromagnetic moments rotate towards the staggered spin polarisation. (b) Clockwise (cw) and counter-clockwise (ccw) rotation of the sublattice magnetisations for reversed N\'eel vector states. Corresponding current dependent longitudinal resistance variation $\pm \delta R_{xx}$ for the N\'eel vector oriented at an angle $\varphi=45^{\circ}$ from the probing current axis.  (c) Current dependent transvers resistance variation  $\pm \delta R_{xy}$ for the N\'eel vector oriented along the probing current axis.
}
\label{figS1}
\end{figure}
\begin{figure}[h!]
\hspace*{1cm}\includegraphics[width=\textwidth]{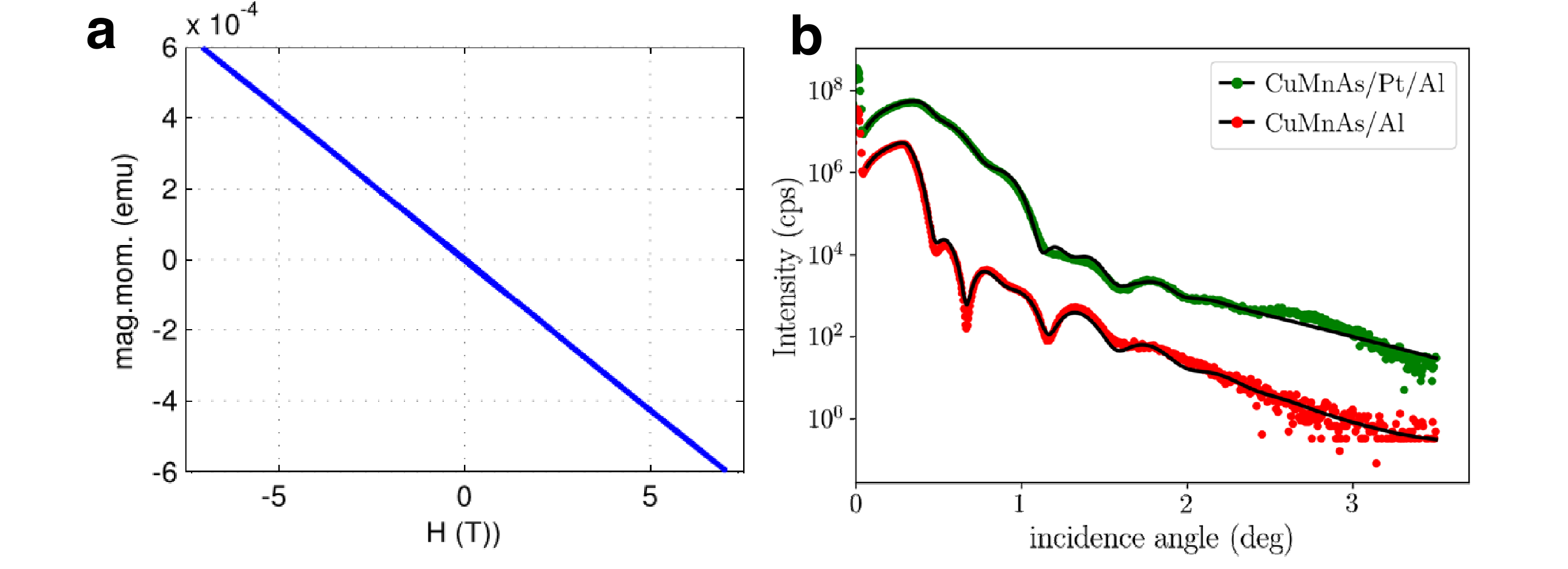}
\caption{(a) SQUID measurement on a CuMnAs(10nm)/Pt(3.6nm)/Al(4nm) film. Magnetization loop up to 7~T shows no indication of a net magnetic moment. (b) X-ray reflectivity measurement on the CuMnAs(10~nm)/Pt(3.6~nm)/Al(4~nm) film. The panel shows the angular dependence of the reflection signal for Cu$\text{K}_\alpha$ radiation at grazing angles. Coloured points show the experimental data while solid lines show our model calculation based on the Parrat formalism. Data for the sample with Pt were scaled by a factor of 10 for clarity.
}
\end{figure}

\begin{figure}[h!]
\hspace*{1cm}\includegraphics[width=.8\textwidth]{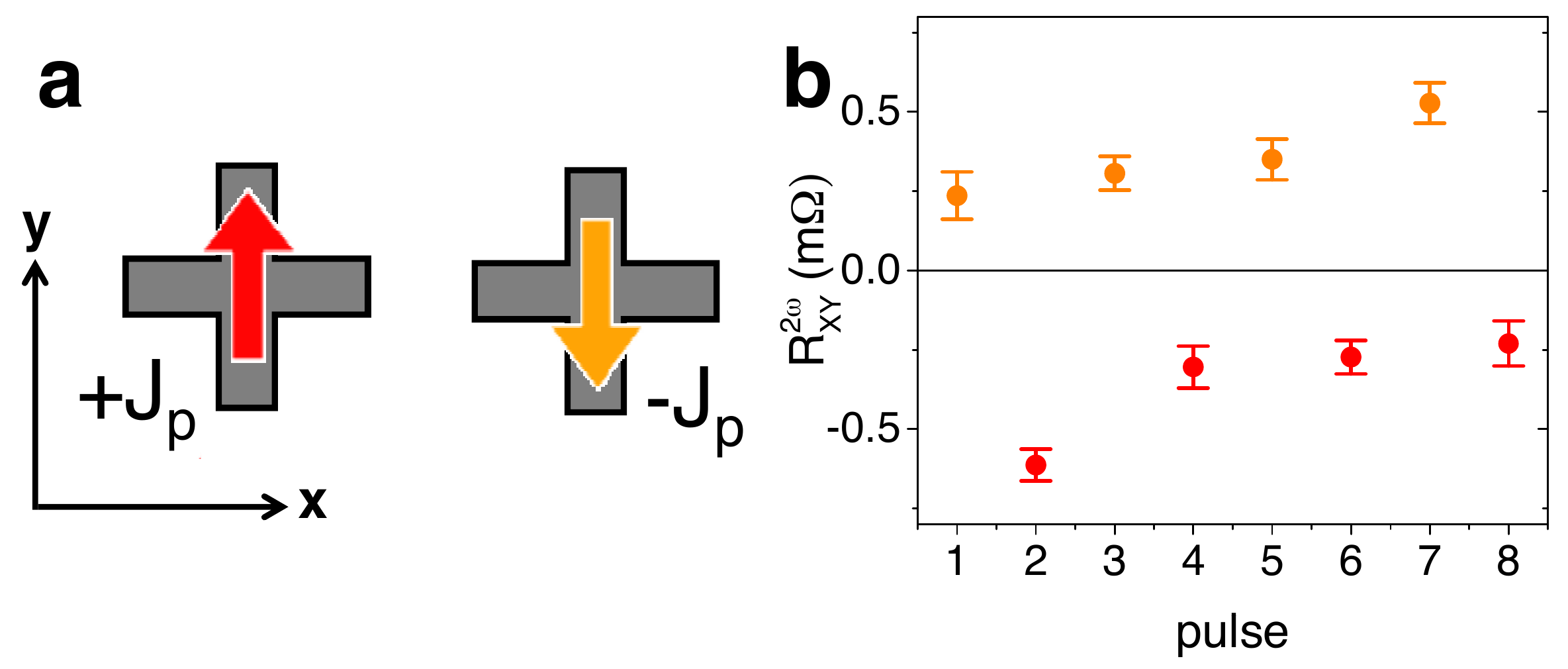}
\caption{(a) Measurement set up with writing pulses along $+y$ (red) and  $-y$ (orange) directions in a $10~\mu$m wide 4 terminal cross bar device patterned from a reference CuMnAs(10nm)/AlOx film without Pt layer. The writing pulse amplitude  $J_p = 9$~mA, the pulse duration $\tau_p = 20$~ms.  (b) Second-harmonic transverse resistance $R_{xy}^{2\omega}$ with probing current of $ J_{ac} =1$~mA. along $x$-axis.
}
\end{figure}
\end{document}